\documentclass[journal=jpcl,manuscript=letter]{achemso}
\usepackage[T1]{fontenc} 
\usepackage{hyperref}
\usepackage{amsmath}
\usepackage{xfrac}

\author{Dominike A. P. de Deus}
\affiliation{Instituto Federal de Educação, Ciência e Tecnologia de Goiás, Departamento de Áreas Acadêmicas, Campus Jataí, Orminda Vieira de Freitas 775, 75804714 Jata\'{\i}, Goi\'as, Brazil}

\author{Daniel  F. Souza}
\affiliation{Instituto Federal do Amazonas, 69020-120, Manaus, AM, Brazil}

\author{Andr\'eia L. da Rosa}
\affiliation{Instituto de F\'isica, Universidade Federal de Goi\'as, Campus Samambaia, 74690-900, Goi\^ania, Goi\'as, Brazil}
\affiliation{Bremen Center for Computational Materials Science, University of Bremen, Am Fallturm 1, 28359, Bremen, Germany}

\author{Renato B. Pontes}
\affiliation{Instituto de F\'isica, Universidade Federal de Goi\'as, Campus Samambaia, 74690-900, Goi\^ania, Goi\'as, Brazil}

\author{Th. Frauenheim}
\affiliation{Bremen Center for Computational Materials Science, University of Bremen, Am Fallturm 1, 28359 Bremen, Germany}
\affiliation{Shenzhen Computational Science and Applied Research Institute, Shenzhen, China}
\affiliation{Beijing Computational Science Research Center, Beijing, China}

\email{andreialuisa@ufg.br}
\phone{+55 62 35211122}

\title{Stability of subnanometer MoS wires under realistic environment}

\abbreviations{xx}

\begin{document}

%
%
%

\begin{abstract}
   We carried out first-principles density functional theory
   calculations of hydrogen and oxygen adsorption and diffusion on subnanometer MoS
   nanowires. The nanowires are robust against adsorption of
   hydrogen. On the other hand, interaction with oxygen shows that the
   nanowires can oxidize with a small barrier. Our results open the path for
   understanding the behavior of MoS nanowires under realistic
   environment.
\end{abstract}


\section{Introduction}

The discovery of graphene has attracted the interest of the scientific
community in other low-dimensional materials, such as transition-metal
dichalcogenides (TMDCs), which exhibit novel physical and chemical
properties with promising applications in
nanotechnology\cite{Chowalla2013,Wang2012,Choi2017}. In particular,
single-layered MoS$_2$ posses a sizebale electronic band gap of
1.8\,eV, which may be useful for applications in high-end electronics,
spintronics, optoelectronics, energy-harvesting, flexible electronics,
and biomedicine\,\cite{NRM2017,PCCP2021,CHEMREV2020}.

Recently transition metal chalcogenide (TMC) subnanometer nanowires
have been fabricated by ion and electron beam techniques in MoS$_2$
layers\,\cite{pantelides,Liu2013}. Contrary to single-layered MoS$_2$,
MoS nanowires show metalic behavior\,\cite{Souza2019}. Such wires
under tensile strain undergo a phase transformation is observed in the
MoS NWs which results in tension-induced hardening in their tensile
modulus\,\cite{CMS2020}. Furthermore, topological states in
one-dimensional MoS nanowires have been proposed\,\cite{Jin2020}.

For several nanoscale applications, the interaction of nanostructures
with the environment is very important, since it provides information
on their stability against degradation. Usually two-dimensional
MoS$_2$ exhibits unique surface effects but the interaction with
hydrogen is limited by the amount of active sites\,\cite{Niu2021}. Although
hydrogen molecules diffuse on MoS$_2$ monolayers with a high energy
barrier of 6.56\,eV, their diffusion can be facilitated by applying
strain which then reduces the diffusion barrier to
0.57\,eV\,\cite{Koh2012}. Reactivity can further be improved via
doping with transition metals\,\cite{Zhao2017}. More recent results
claim that oxygen strongly interacts with defective MoS$_2$ layers,
enhancing their photoluminescence properties due to dissociation of
O$_2$ on sulphur vacancies\,\cite{JPCC2021}.

Although the interaction of MoS$_2$ monolayers with oxygen and
hydrogen has been largely explored, investigations of MoS nanowires
under realistic environments are still missing.  In this paper we
perfrom first-principles calculations density functional theory
calculations of hydrogen and oxygen adsorption and difusion on
subnanometer MoS nanowires. We show that the nanowires have sizeable
interaction with atomic hydrogen. However, dissociation of hydrogen
molecules is hindered by a high enegy barrier and unfavourable
orientation of orbitals. On the other hand, oxidation of MoS wires may
take place with a small diffusion barrier of oxygen atoms.

\section{Computational details}

The calculations are performed using density-functional theory (DFT)\,\cite{Hohenberg:64,Kohn:65},
as implemented in the Vienna ab initio simulation package (VASP)\,\cite{Kresse:99}, The
electron-ion interactions are taken into account using the Projector
Augmented Wave (PAW) method.  We employ the generalized gradient
approximation (GGA-PBE)\,\cite{Perdew:96} to describe the interaction between valence
electrons. The Kohn-Sham orbitals are expanded in a plane wave basis
set with an energy cutoff of 500\,eV. The Brillouin Zone is sampled
according to the Monkhorst-Pack method using a (1$\times$1$\times$8)
{\bf k}-points mesh. In order to determine the diffusion and reaction
barriers we have performed CI-NEB\,\cite{Henkelman1,Henkelman2} calculations with
seven images to search the minimum-energy reaction paths and saddle
points between the initial state and final state configurations.

\section{Results and discussion}

\subsection{Pristine nanowires}

\begin{figure}[!htp]
 \centering
  \includegraphics[width=8cm, keepaspectratio=true, clip=true]{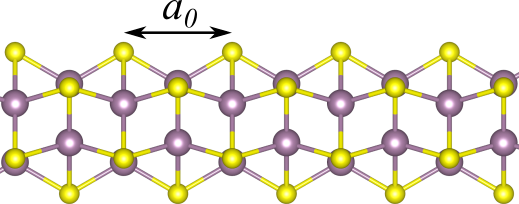}\\
 \includegraphics[width=8cm, keepaspectratio=true, clip=true]{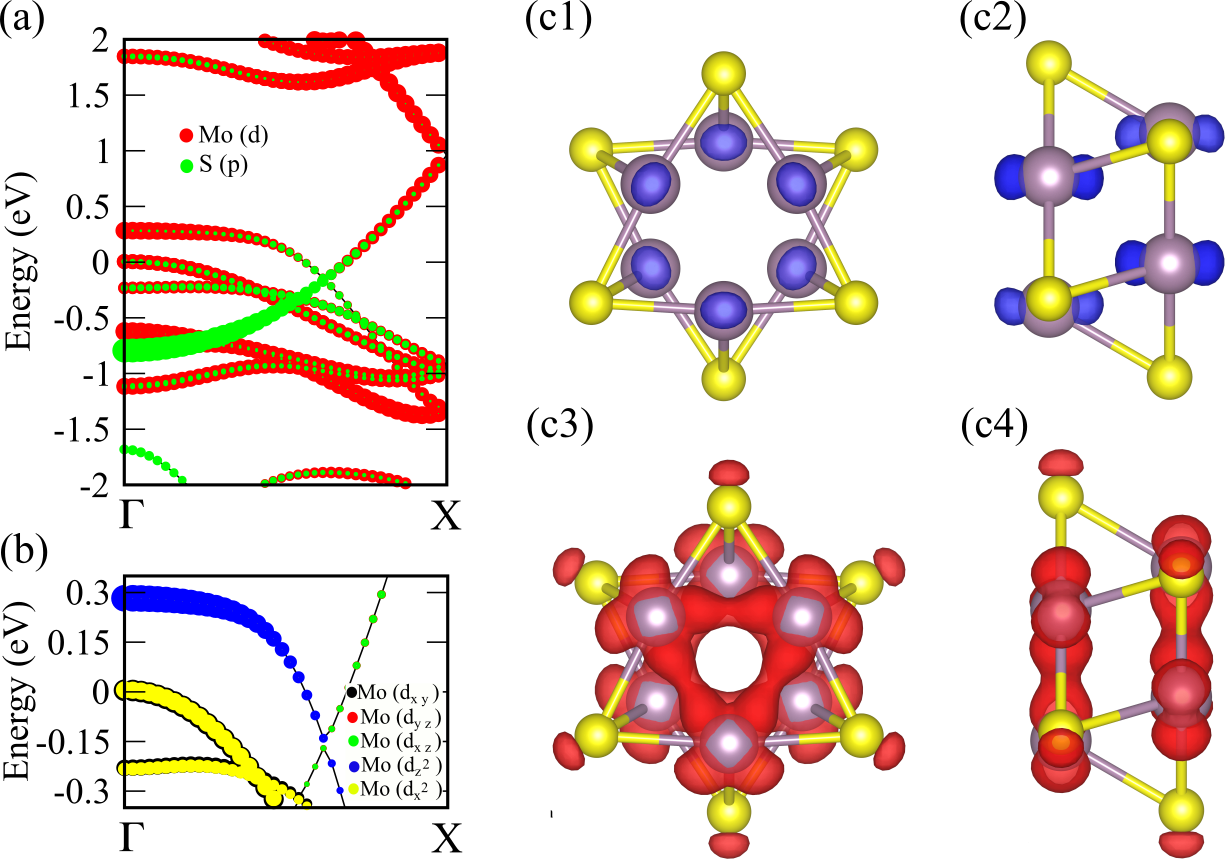}
  \caption{\label{fig:structure_pristine} Relaxed geometry of an isolated MoS nanowire with lattice parameter $a_{\rm 0}$ along the wire growth direction. Electronic band structure of pristine MoS nanowire. a) Orbital decomposed band structure, b) Orbital Mo-$d$
     decomposed band structure and c1)-c4) band projected charge density. Blue are empty states (0.4 eV above the Fermi level) and red are
     occupied states (0.4 eV below the Fermi level, which is set at zero).}
\end{figure}

Mo and S atoms in pristine MoS nanowires are arranged by stacking of
triangular layers rotated 180$^{\rm o}$ along the nanowire axial
direction, with three S atoms located at the vertices of the triangles
and three Mo atoms located between the S atoms, as seen in
Fig.\,\ref{fig:structure_pristine}. The calculated equilibrium lattice
parameter along the axis wire is 4.39\,{\AA}. Contrary to
single-layered MoS$_2$, pristine MoS nanowires show metallic behavior,
as seen in Fig.\,\ref{fig:structure_pristine}(a). At $\Gamma$ point
the states with character $d_{\rm x^2}$ cross the Fermi
level. Localized states with $d_{\rm z^2}$ character lie around 0.25
eV above the Fermi level, as seen in
Fig.\,\ref{fig:structure_pristine}(b). In
Fig.\,\ref{fig:structure_pristine}(c1) and (c2) (empty states) and
(c3) and (c4) (occupied states) we show the orbital projected charge
density in pristine wires. The empty states lie along
the wire growth direction.  We discuss the importance of orbital
orientation later on.

In order to understand the strength between MoS wires and hydrogen/oxygen we have calculated the adsorption energy according to:

\begin{equation}
E_{\rm ads} = E_{\rm MoS-ad} - E_{\rm MoS} - \sum_i{N_i\mu_i},
\end{equation}

where E$_{\rm MoS-ad}$ and E$_{\rm MoS}$ are the total energies of the MoS
supercell with and without the adsorbate (hydrogen or oxygen). N$_i$ and $\mu_i$ are the number and chemical potencial of hydrogen and oxygen in gas phase. 

\begin{table}
  \caption{Adsorption energy E$_{\rm ads}$ and diffusion/dissociation barrier  $\Delta{\rm E}$ of hydrogen and oxygen on MoS nanowires.}
 \label{table:formation}
 \begin{tabular*}{8cm}{@{\extracolsep{\fill}}lccc}
     \hline
adsorbate & $\%$ (ML) & E$_{\rm ads}$ (eV) & $\Delta{\rm E}$ (eV) \\
\hline
H on S  & $\sfrac{1}{12}$ & -0.55    &    \\
H on Mo & $\sfrac{1}{12}$ & -0.30    &  \\
H on S  & $\sfrac{1}{24}$ & -0.80    &  \\
H on Mo & $\sfrac{1}{24}$ & -0.60    & \\
\hline
H$_2$   & $\sfrac{1}{12}$ & -0.006   & \\
H$_2$   & $\sfrac{1}{24}$ & -0.005   &  \\
\hline
H       & $\sfrac{1}{24}$ &          & 0.35 \\
H$_2$   & $\sfrac{1}{12}$ &          & 2.60 \\
H$_2$   & $\sfrac{1}{24}$ &          & 1.10\\
\hline
O on S  & $\sfrac{1}{12}$ &  0.58    & \\
O on Mo & $\sfrac{1}{12}$ & -0.22    & \\
O on S  & $\sfrac{1}{24}$ & -0.19    &\\
O on Mo & $\sfrac{1}{24}$ & -0.50    &\\
\hline
O$_2$   & $\sfrac{1}{12}$ &  0.09    &\\
O$_2$   & $\sfrac{1}{24}$ & -0.25    &\\
\hline
O       & $\sfrac{1}{24}$ &          &  0.42\\
O$_2$   & $\sfrac{1}{24}$ &          & 0.20\\
\hline
\end{tabular*}
\end{table}

\begin{figure}[!htp]
 \centering
 \includegraphics[width=8cm, keepaspectratio=true, clip=true]{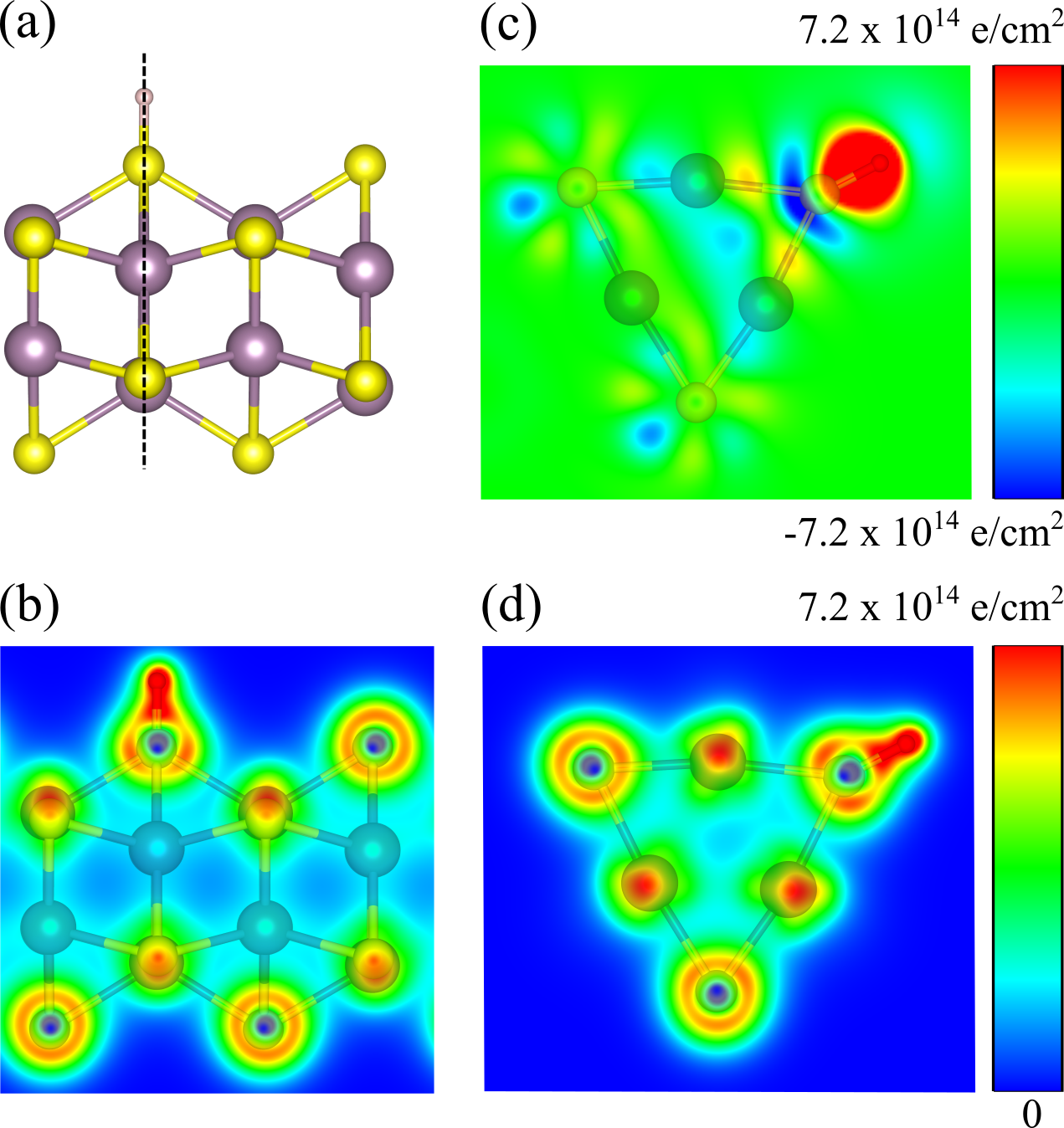}
 \caption{\label{fig:charge112hatom} a) Relaxed structure of atomic hydrogen on MoS wires, b) charge difference $\rho$, c) total charge density (cut parallel) and d) cut perpendicular to the growth direction. Coverage of $\sfrac{1}{24}$\,ML.}
 \end{figure}

We have considered adsorption of atomic and molecular
hydrogen/oxygen. Concentrations of $\sfrac{1}{12}$ and $\sfrac{1}{24}$
monolayers (ML) which correspond to $(1\times1\times2)$ and
$(1\times1\times1)$ unit cells, respectively, have been used. The
results for the adsorption energy are shown in Table\,\ref{table:formation}.  For atomic
hydrogen, coverages of $\sfrac{1}{12}$\,ML shows that adsorption of
atomic hydrogen atom has the lowest energy on a sulphur atom.  This energy
is higher by 0.25\,eV than adsorption on a Mo atom. Smaller coverages
of $\sfrac{1}{24}$\,ML have a similar behavior. For hydrogen concentration of
$\sfrac{1}{24}$\,ML, S-H bond length is $d_{\rm S-H}$=
1.36\,{\AA}. The relaxed structure is shown in
Fig.\,\ref{fig:charge112hatom}(a). Atomic hydrogen prefers to be at ontop position, with the nanowire structure
barely affected by the hydrogen atom. Charge density difference
calculated according to $\Delta\rho = \rho_{\rm MoS-H} - \rho_{\rm MoS} - \rho_{\rm H}$ is shown in
Fig.\,\ref{fig:charge112hatom}(b) with the nanowire structure
barely affected by the hydrogen atom. Total charge density with views along and perpendicular to the wire growth direction is shown
in Fig.\,\ref{fig:charge112hatom}(c) and (d). 

\begin{figure}[!htp]
 \centering
 \includegraphics[width=8cm, keepaspectratio=true, clip=true]{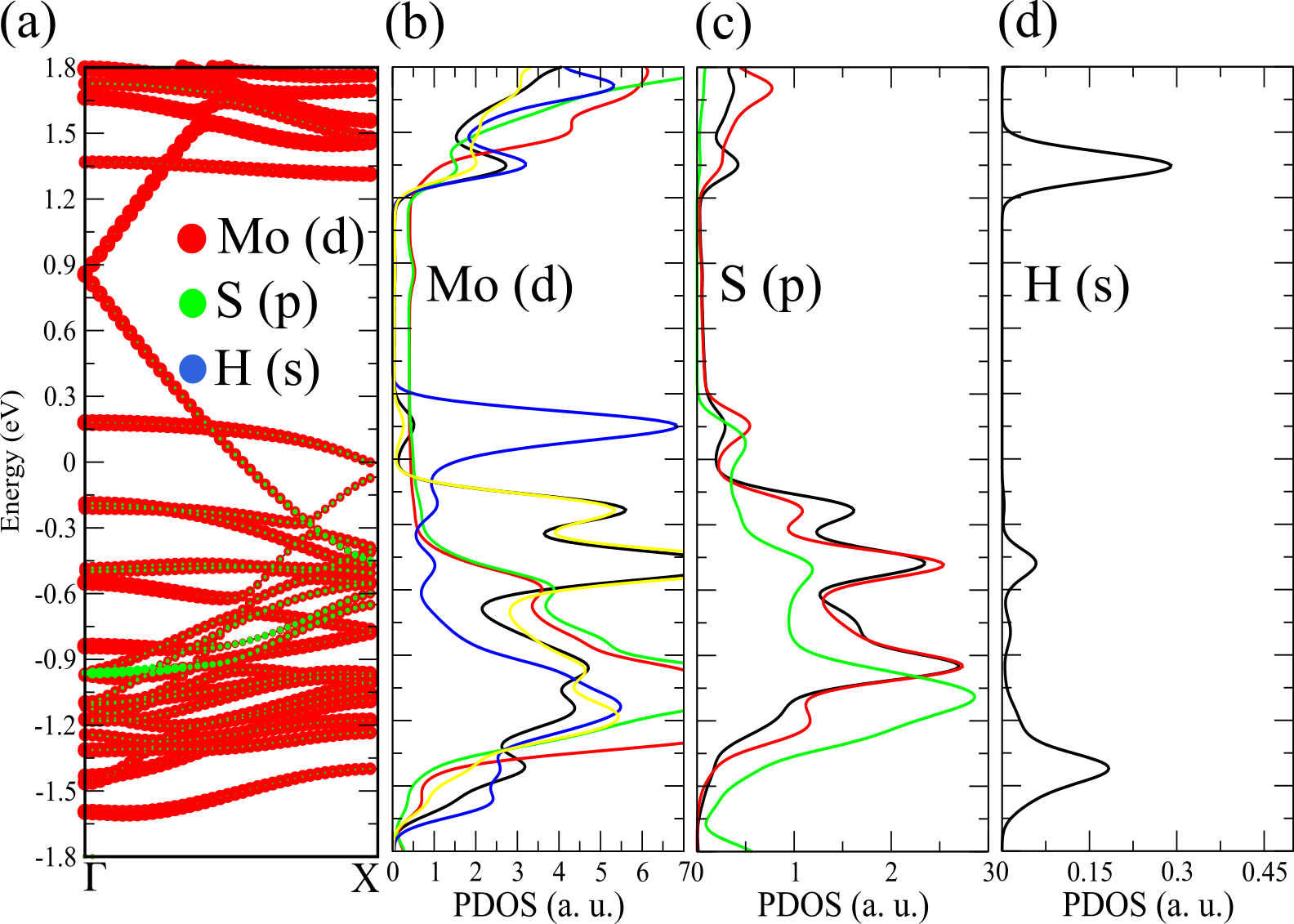}
    \caption{\label{fig:band112hatom} Band structure of atomic hydrogen adsorbed on MoS nanowire for $\sfrac{1}{24}$\,ML coverage. a) orbital projected band structure, b) PDOS on Mo-d, c) PDOS on S-p and d) PDOS on H-s.}
 \end{figure}

To corroborate with this picture we show the band structure of atomic
hydrogen adsorbed on MoS nanowire for $\sfrac{1}{24}$\,ML coverage. a)
orbital pr\ ojected band structure, b) PDOS on Mo-d, c) PDOS on S-p
and d) PDOS on H-s in Fig.\ref{fig:band112hatom}.

\begin{figure*}[!htp]
 \centering \includegraphics[width=\textwidth, keepaspectratio=true,clip=true]{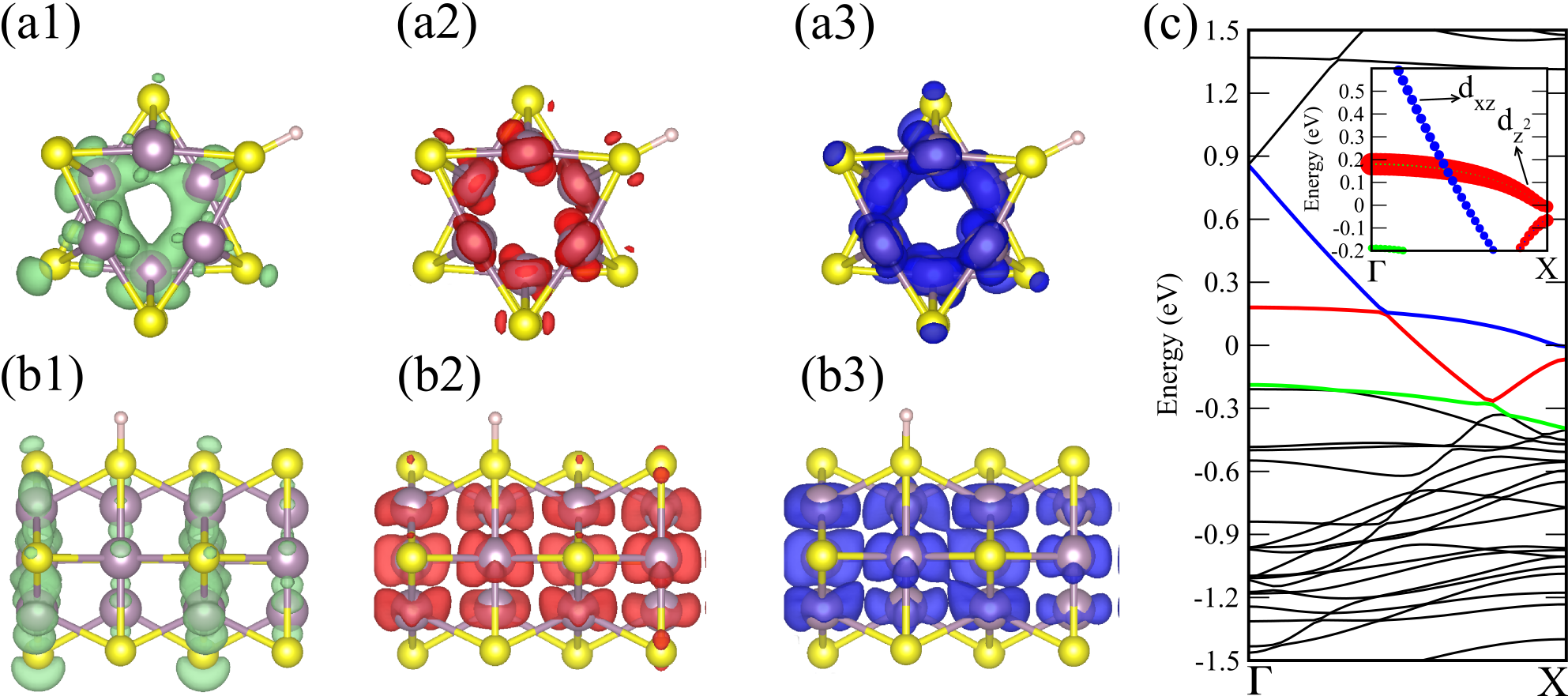}
 \caption{\label{fig:pcharge112hatom}  Orbital projected charge density of atomic hydrogen on MoS nanowire for $\sfrac{1}{24}$\,ML coverage. a1) and b1)  Mo-$d_{\rm xy}$, a2) and b2) Mo-$d_{\rm xz}$, a3) and b3)  Mo-$d_{\rm z^2}$ and c) projected band structure corresponding to states shown in a) and b). The Fermi level is set at zero.}
\end{figure*}

In order to further understand this behavior we have calculated the
corresponding orbital projected band structure. The states comes from
Mo $d_{\rm z^2}$ (red) and $d_{\rm xz}$ (blue) and Mo $d_{\rm xy}$
(green) orbitals.  Fig.\,\ref{fig:pcharge112hatom} reveals states with
$d_{\rm z^2}$ character around the Fermi level which are shifted
downwards in energy compared to the ones in pristine wires. The H-S
bond has major contributions for the PDOS at -0.4\,eV in
Figs.\,\ref{fig:pcharge112hatom}(b)-(d).

Orbital projected charge density of atomic hydrogen on MoS nanowire at
$\sfrac{1}{24}$\,ML coverage. Close to the Fermi level we identified
Mo-$d_{\rm xy}$ (a1) and (b1), Mo-$d_{\rm xz}$ (a2) and (b2) and
Mo-$d_{\rm z^2}$ (a3) and (b3). In c) we show the projected band
structure corresponding to these states.

\begin{figure}[!htp]
 \centering
 \includegraphics[width=8cm, keepaspectratio=true, clip=true]{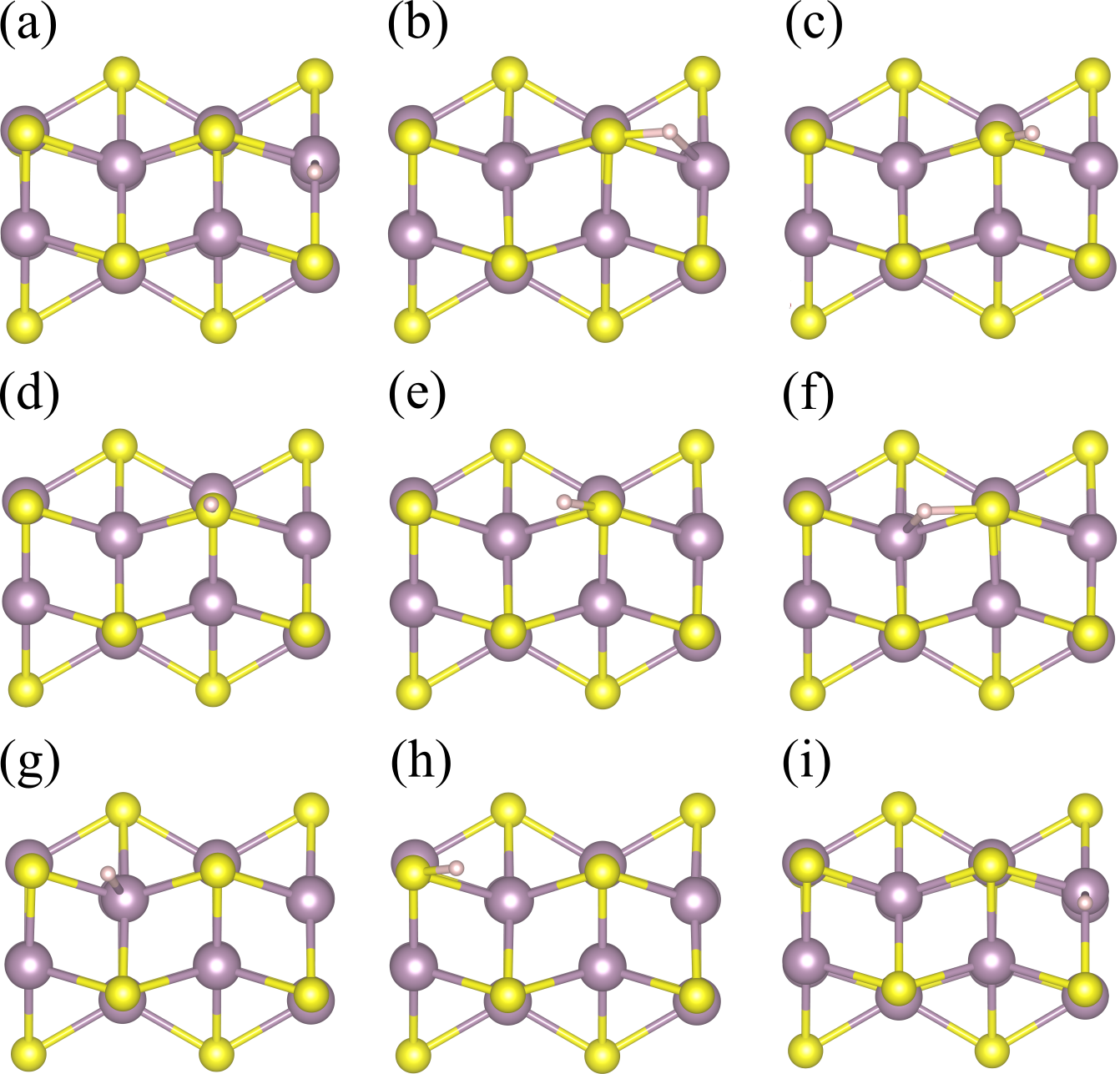}\\\vspace{1cm}
  \includegraphics[width=8cm, keepaspectratio=true, clip=true]{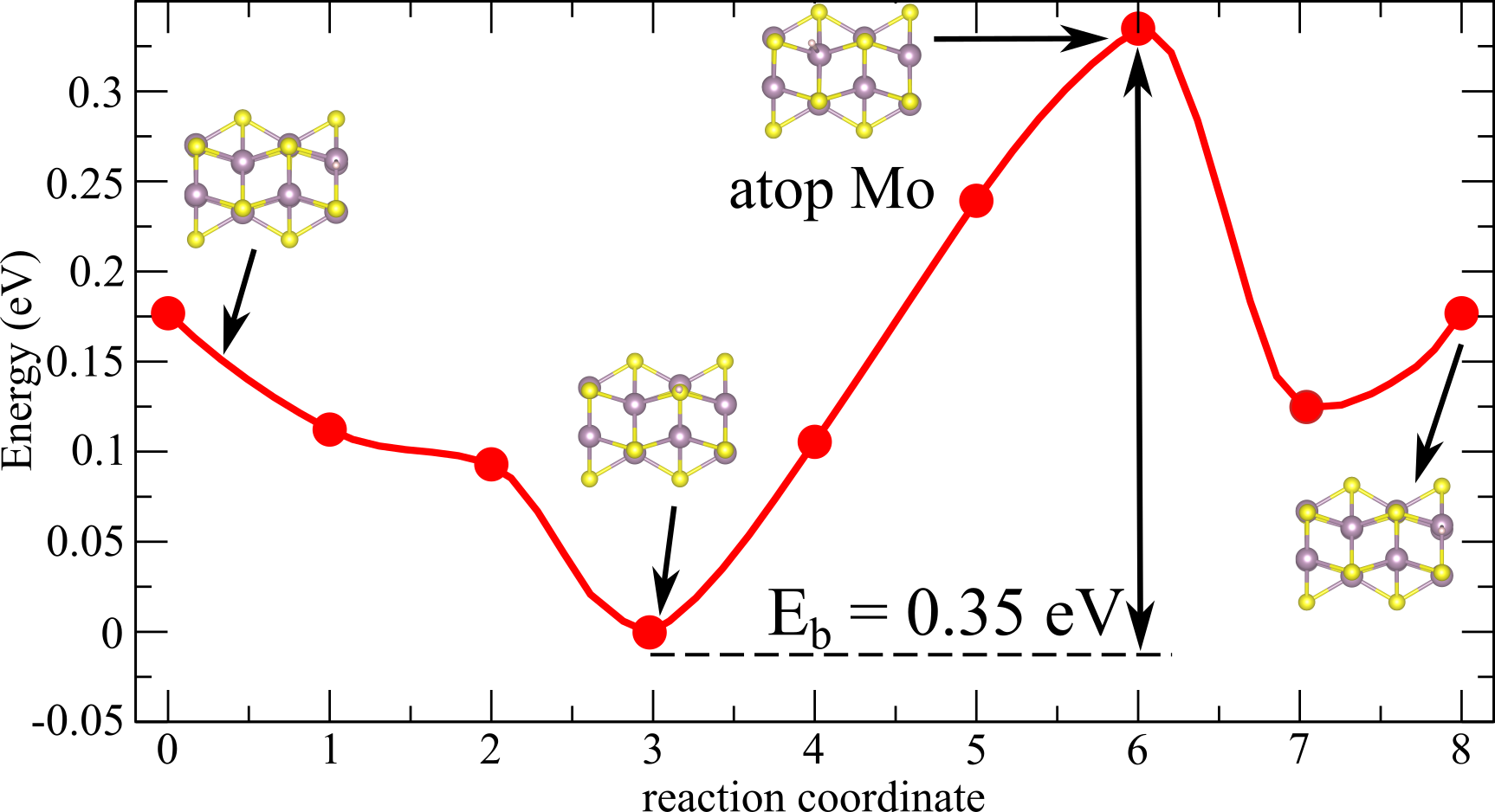}
    \caption{\label{fig:path112hatom} a)- i) Minimum-energy reaction path and j) diffusion barrier for atomic hydrogen on MoS nanowire for $\sfrac{1}{24}$\,ML coverage.}
 \end{figure}

The transition state along the considered path shown in
Fig.\,\ref{fig:path112hatom} has the hydrogen adsorbed at the ontop
position of a Mo atom. The dissociation barrier is 0.35\,eV, as it can
be seen in Fig.\ref{fig:path112hatom} (j).  By increasing the hydrogen
coverage to $\sfrac{1}{12}$\,ML the barrier is 2.60\,eV (not
shown). As a matter of comparison, usual barriers for atomic hydrogen
are usually a few meV 0.05-0.3\,eV, on hcp(0001), fcc(111) and
bcc(110) metal surfaces\,\cite{SurfSci2012,PCCP2021}, which is
consistent with our results at low coverages.

Table 1 shows that contrary to atomic hydrogen, the adsorption energy
of molecular hydrogen is only 6.00 meV. Furthermore, negligible
difference is found for hydrogen concentrations of $\sfrac{1}{24}$ and
$\sfrac{1}{12}$\,ML. In order to further understand this behavior, we
have calculated the dissociation barrier of hydrogen along the path
shown in Fig.\,\ref{fig:path112hatom}(a)-(i). Barrier of 1.10 eV.

\begin{figure}[!htp]
\centering
\includegraphics[width=8cm, keepaspectratio=true, clip=true]{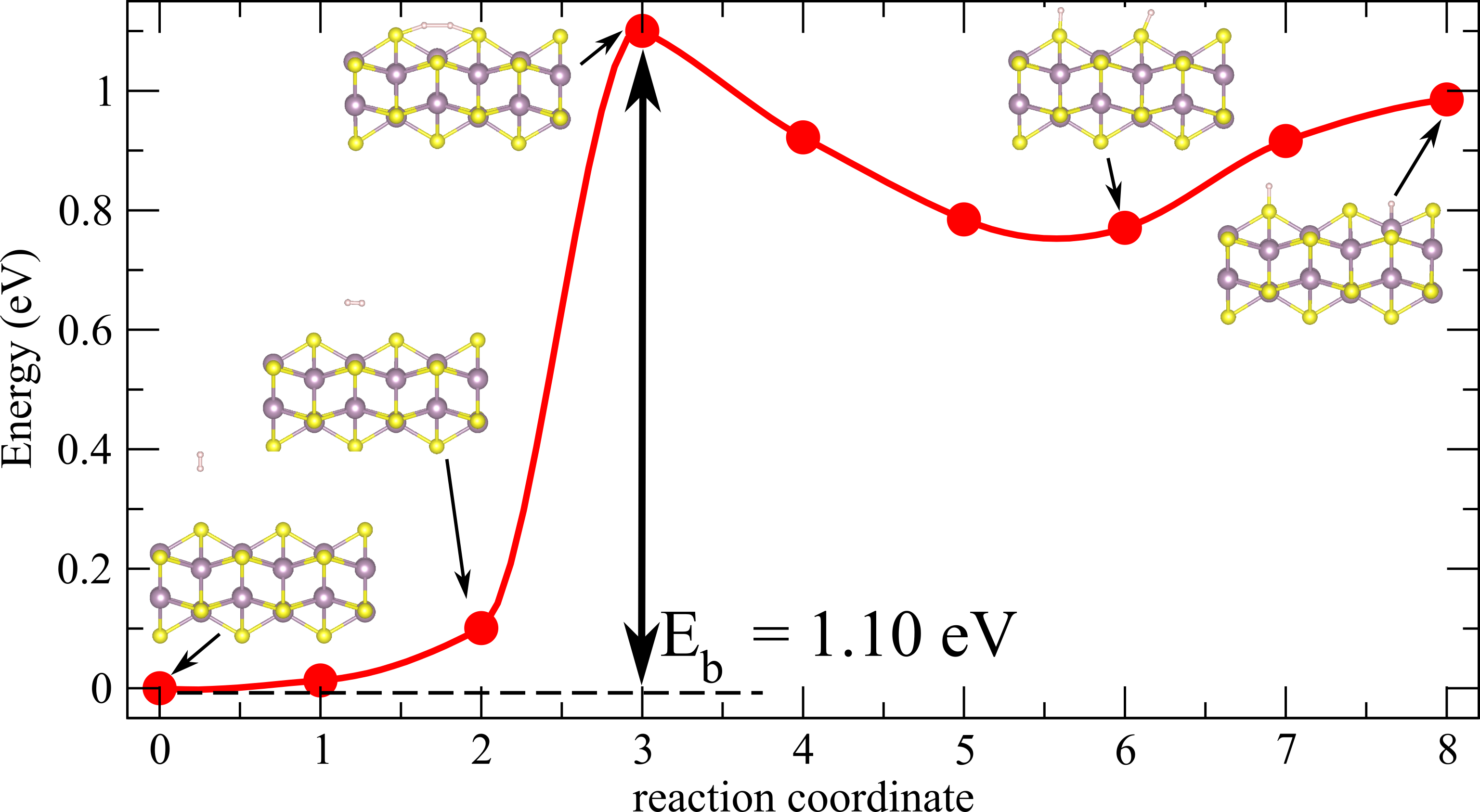}\\
\includegraphics[width=8cm, keepaspectratio=true, clip=true]{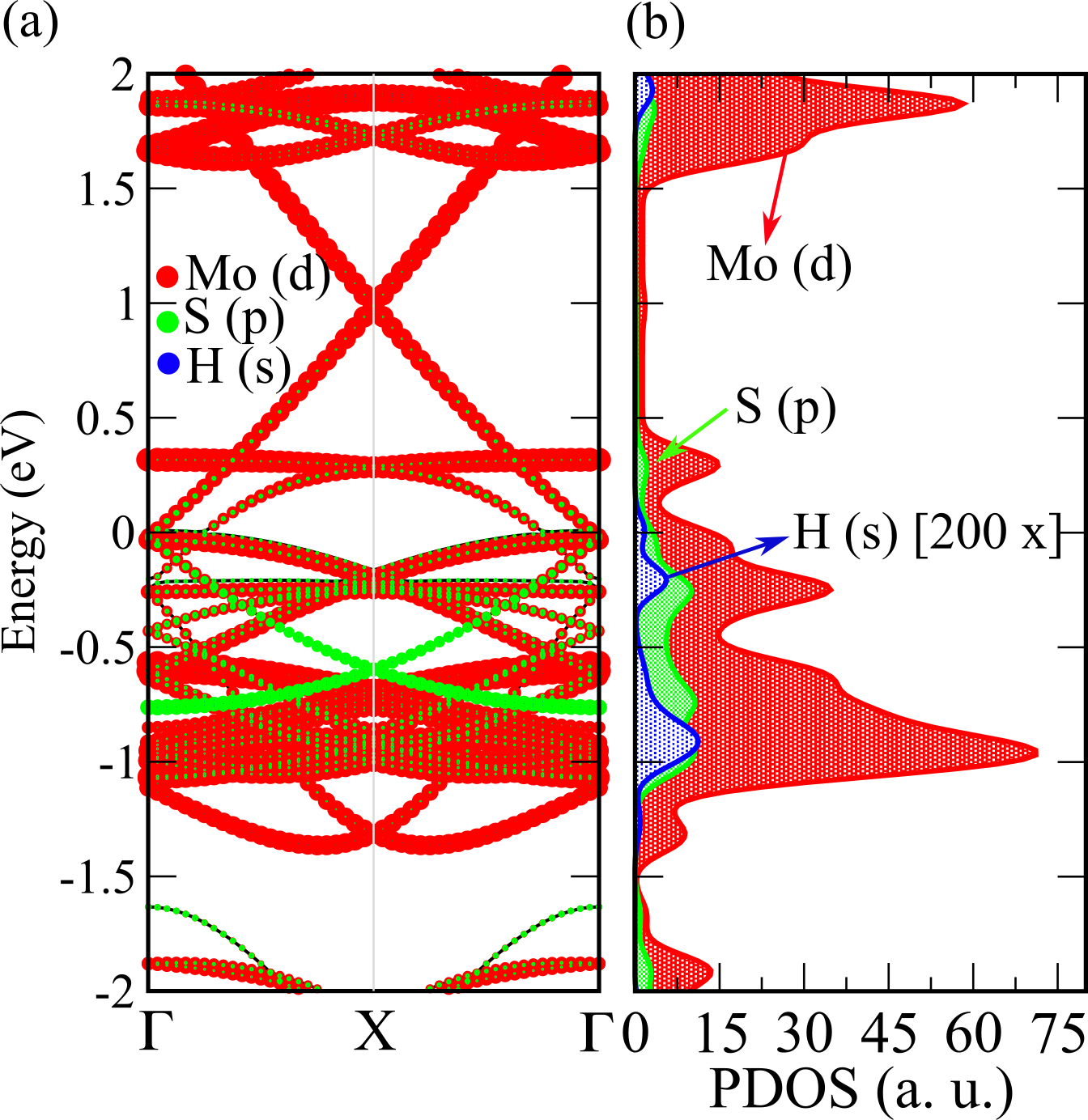}
 \caption{\label{fig:neb112h2} Dissociation barrier of a hydrogen molecule on MoS nanowire for $\sfrac{1}{24}$\,ML
      coverage.  Band structure and PDOS corresponding to the most stable state shown a).}
\end{figure}

The band structure of the most stable state in Fig.\,\ref{fig:neb112h2}(a) is shown in Fig.\,\ref{fig:neb112h2}(b).

\subsubsection{Oxygen adsorption}

\begin{figure}[!htp]
 \centering
 \includegraphics[width=8cm, keepaspectratio=true, clip=true]{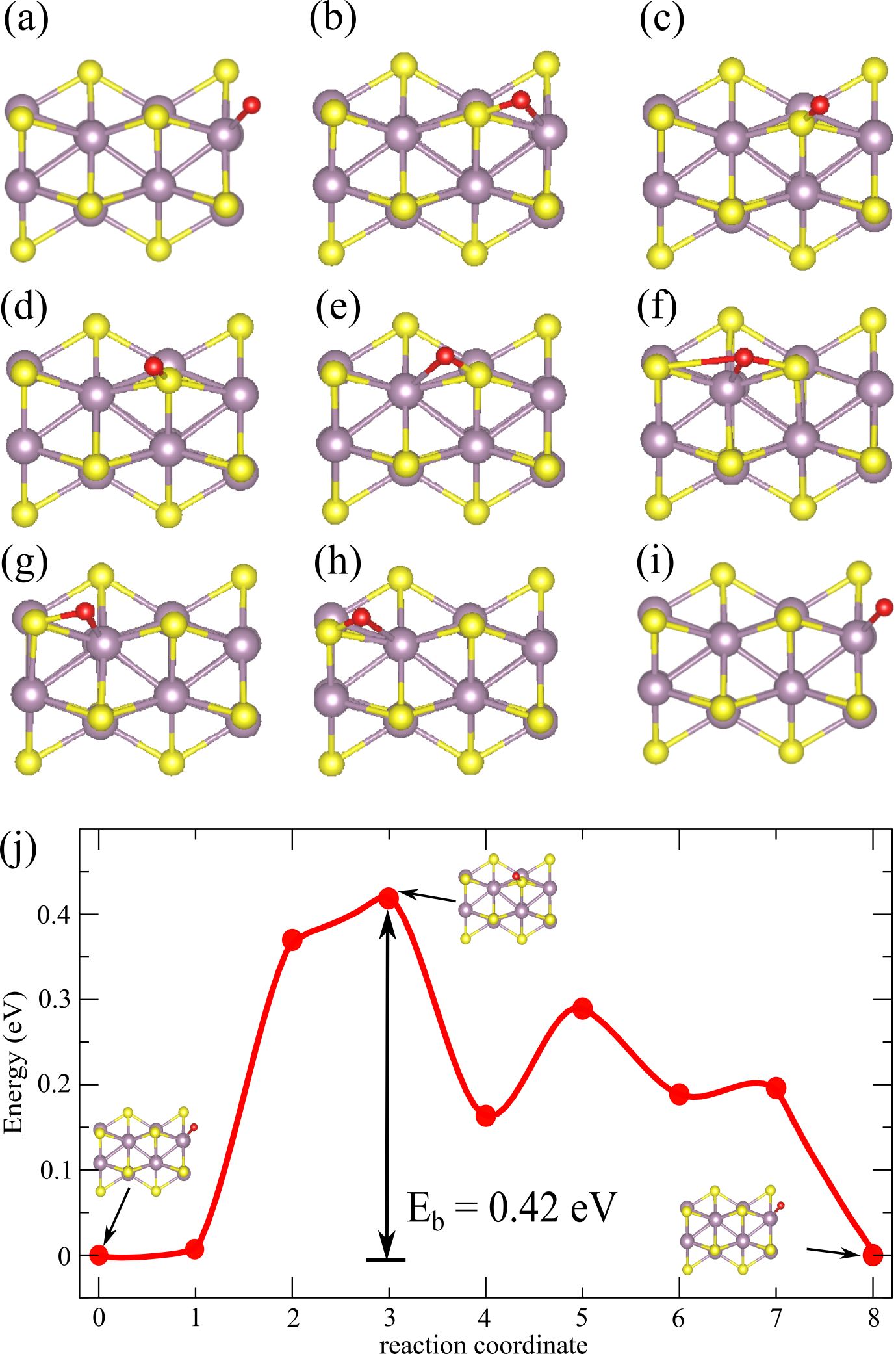}
    \caption{\label{fig:path112oatom} Minimum-energy reaction path for atomic oxygen atom on MoS nanowire with hydrogen adsorbed on Mo-S bridge for $\sfrac{1}{24}$\,ML coverage.}
\end{figure}

\begin{figure}[!htp]
\centering
 \includegraphics[width=8cm, keepaspectratio=true, clip=true]{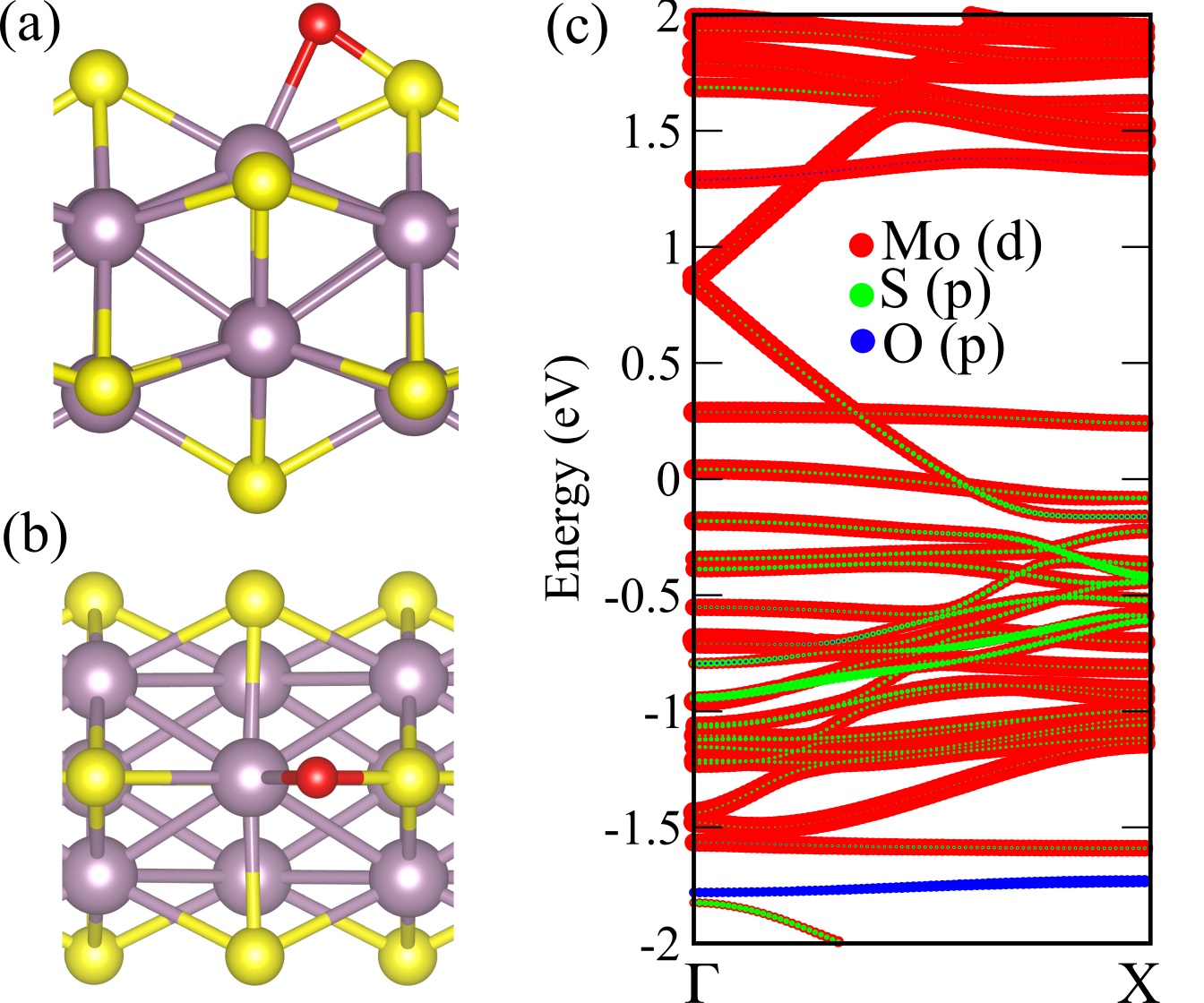}
 \caption{\label{fig:model-banda-mos-o} Band structure of atomic oxygen on MoS for the lowest energy structure shown in Fig.\,\ref{fig:path112oatom}.}
\end{figure}

Atomic oxygen atom adsorbs at a bridge position as shown in Fig.\ref{fig:path112oatom}(a). Starting from this
configuration, Fig.\,\ref{fig:path112oatom}(a)-(i) show the minimum
energy path with a diffusion barrier of 0.42\,eV. The transition state
corresponds to the oxygen atom close to an ontop position. The band structure corresponding to Fig.\,\ref{fig:path112oatom}(a)) is shown in Fig.\,\ref{fig:model-banda-mos-o}(j).

At $\sfrac{1}{12}$\,ML oxygen concentration as shown in Fig.\,\ref{fig:model-banda-mos-o} a) side view) and b) top view we have an endothermic reaction, as seen from Table\,\ref{table:formation}. Interestingly enough, by decreasing the oxygen concentration to $\sfrac{1}{24}$\,ML, the reaction turns out to be exothermic.  The band structure is shown in
Fig.\,\ref{fig:model_banda_pdos_mos112_o2_estavel}(c). The more
stable structure for an oxygen molecule shows a small magnetic moment due to
outwards relaxation of nearby Mo atoms. The presence of the O$_2$ molecule near a Mo atom
introduces O-$p_{\rm z}$ states close to the $\Gamma$ and X
points. Furthermore, the valence band maximum (VBM) has mainly
Mo-$d_{\rm z^2}$ character, whereas the conduction band minimum (CBM) shows
mixed O-$p_{\rm z}$, Mo-$d_{\rm xz}$ and Mo-$d_{\rm yz}$ states, as
seen in Fig.\,\ref{fig:model_banda_pdos_mos112_o2_estavel}(d).

\begin{figure}[!htp]
\centering
 \includegraphics[width=8cm, keepaspectratio=true, clip=true]{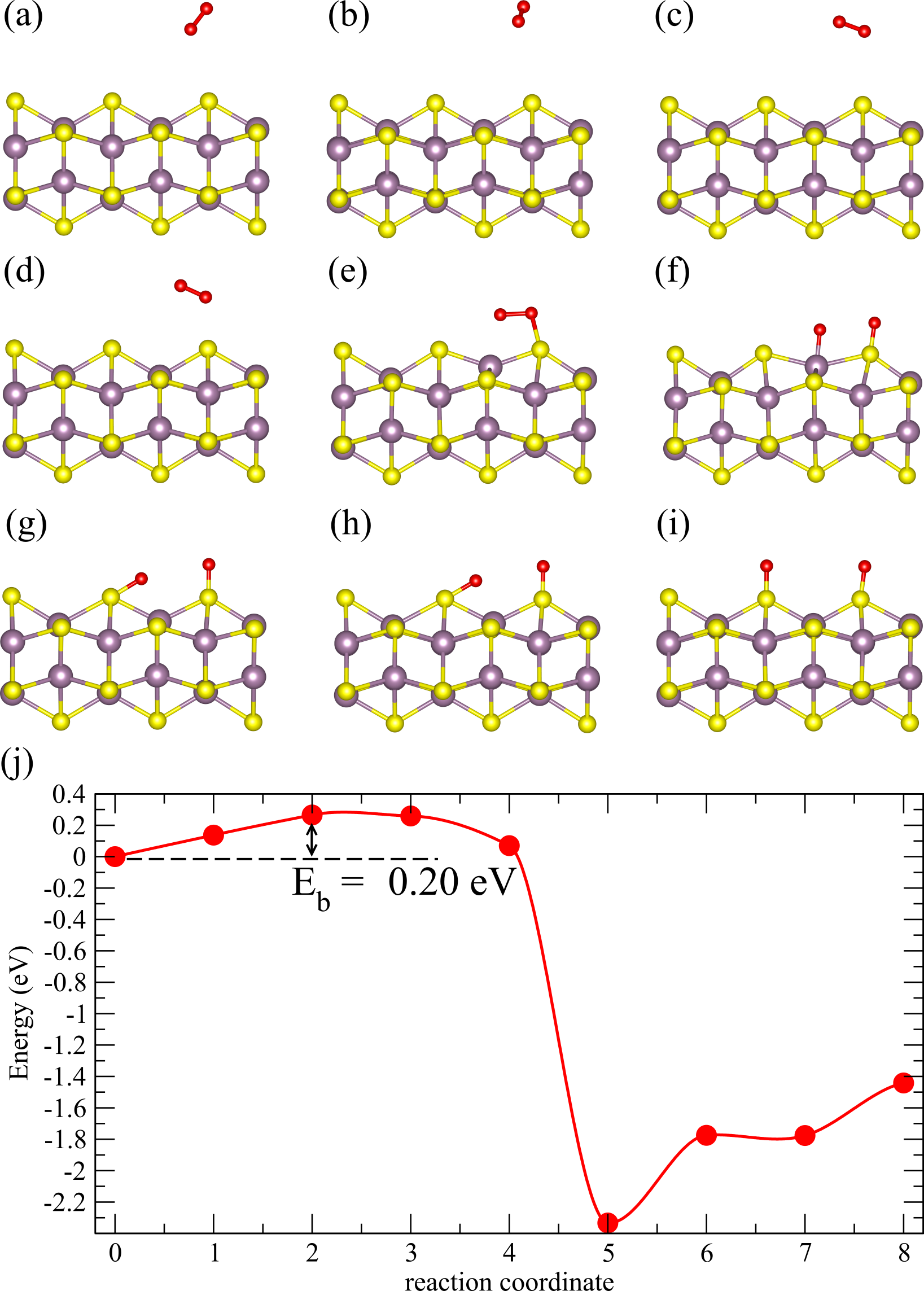}
 \caption{\label{fig:neb_112_o2} Minimum-energy reaction path for molecular oxygen on MoS nanowire at bridge position  at $\sfrac{1}{12}$\,ML coverage.}
\end{figure}

\begin{figure}[!htp]
\centering
 \includegraphics[width=8cm, keepaspectratio=true, clip=true]{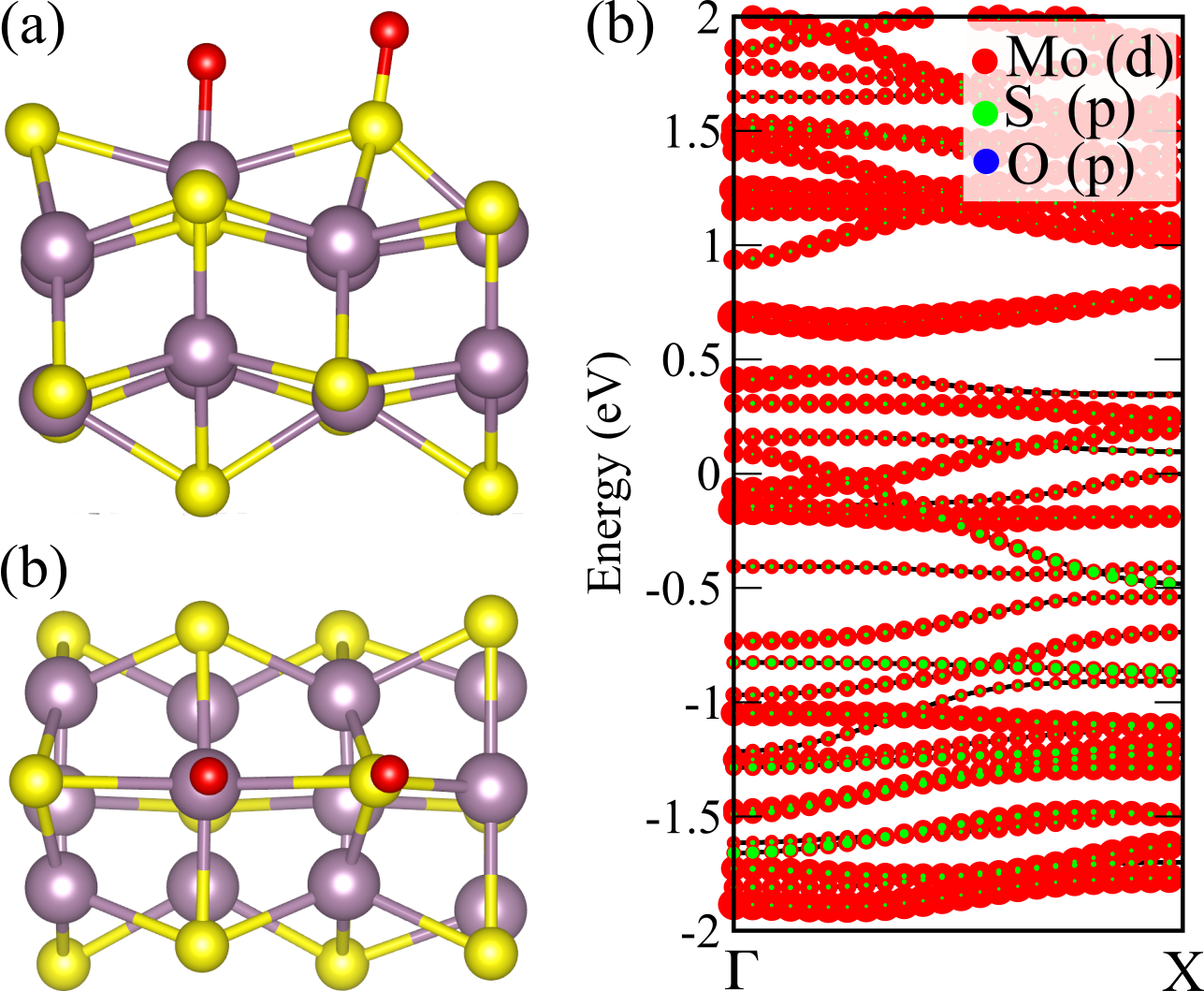}
 \caption{\label{fig:model_banda_pdos_mos112_o2_estavel} Electronic band structure and PDOS of the most stable configuration shown in Fig.\,\ref{fig:neb_112_o2}.}
\end{figure}

\section{Conclusion}

In this paper we perfrom first-principles calculations density
functional theory calculations of hydrogen and oxygen adsorption and
difusion on subnanometer MoS nanowires. We show that the nanowires
have sizeable interaction with atomic hydrogen, but little with
hydrogen molecules. Furthemore, oxidation of MoS nanowires is feasible
and occurs with a small barrier.

\begin{acknowledgement}

We acknowledge the financial support from the Brazilian Agency CNPq
under grant number 313081/2017-4 and 305335/2020-0 and German Science Foundation (DFG) under the
program FOR1616. The calculations have been performed using the
computational facilities of Supercomputer Santos Dumont at LNCC, 
QM3 cluster at the Bremen Center for Computational Materials Science
CENAPAD-SP at Unicamp and  LaMCAD at UFG.

\end{acknowledgement}

%
%

\bibliography{references.bib}


\providecommand{\latin}[1]{#1}
\makeatletter
\providecommand{\doi}
  {\begingroup\let\do\@makeother\dospecials
  \catcode`\{=1 \catcode`\}=2 \doi@aux}
\providecommand{\doi@aux}[1]{\endgroup\texttt{#1}}
\makeatother
\providecommand*\mcitethebibliography{\thebibliography}
\csname @ifundefined\endcsname{endmcitethebibliography}
  {\let\endmcitethebibliography\endthebibliography}{}

\end{document}